\documentclass[sigconf]{acmart}
\renewcommand\footnotetextcopyrightpermission[1]{} 
\usepackage{booktabs} 

\usepackage{algorithm} 
\usepackage{algorithmic} 
\usepackage{mdwlist} 
\usepackage[mathscr]{euscript} 
\usepackage{amsbsy} 
\usepackage{amsfonts} 
\usepackage{subfig} 
\usepackage{array} 
\usepackage{ragged2e} 

\usepackage{graphicx}
\usepackage{amsmath} 
\usepackage{hyperref} 
\usepackage{multirow} 
\usepackage[bottom]{footmisc}
\usepackage{tablefootnote}[2014/01/26] 
\usepackage{color}
\usepackage{xspace}

\usepackage{url}

\setcopyright{none}

\newcommand{\T}[1]{\boldsymbol{\mathscr{#1}}}

\newcommand{\bit}{\begin{itemize*}}
	\newcommand{\eit}{\end{itemize*}}

\newcommand{\hide}[1]{}

\newcolumntype{C}[1]{>{\centering\arraybackslash}p{#1}}
\newcolumntype{R}[1]{>{\RaggedLeft\arraybackslash}p{#1}}
\newcolumntype{L}[1]{>{\RaggedRight\arraybackslash}p{#1}}

\newcommand{\gift}{\textsc{GIFT}\xspace}
\newcommand{\ptf}{\textsc{P-Tucker}\xspace}
\newcommand{\stf}{\textsc{Silenced-TF}\xspace}
\newcommand{\mat}[1]{\mathbf{#1}}
\newcommand{\vect}[1]{\mathbf{#1}}

\newcommand{\algorithmicdoinparallel}{\textbf{do in parallel}}
\makeatletter
\AtBeginEnvironment{algorithmic}{%
	\newcommand{\FORP}[2][default]{\ALC@it\algorithmicfor\ #2\ %
		\algorithmicdoinparallel\ALC@com{#1}\begin{ALC@for}}%
	}
	\makeatother

\begin{document}
\title{GIFT: Guided and Interpretable Factorization for Tensors - 
	\\ An Application to Large-Scale Multi-platform Cancer Analysis}

\renewcommand{\thefootnote}{\fnsymbol{footnote}}
\renewcommand{\footnotesize}{\large} 

\author{Sejoon Oh\footnotemark[1]}
\affiliation{%
	\institution{Seoul National University}
}
\email{ohhenrie@snu.ac.kr}

\author{Jungwoo Lee\footnotemark[1]}
\affiliation{%
	\institution{Seoul National University}
}
\email{muon9401@gmail.com}

\author{Lee Sael}
\affiliation{%
	\institution{The State University of New York (SUNY) Korea}
}
\email{sael@sunykorea.ac.kr}

\begin{abstract}
    \textbf{Motivation:} Given multi-platform genome data with prior knowledge of functional gene sets, how can we extract interpretable latent relationships between patients and genes?
More specifically, how can we devise a tensor factorization method which produces an interpretable gene factor matrix based on gene set information while maintaining the decomposition quality and speed? \\
\textbf{Method:} We propose  \textbf{\gift}, a \textbf{G}uided and  \textbf{I}nterpretable  \textbf{F}actorization for  \textbf{T}ensors.
\gift provides interpretable factor matrices by encoding prior knowledge as a regularization term in its objective function. \\
\textbf{Results:}
Experiment results demonstrate that \gift produces interpretable factorizations with high scalability and accuracy, while other methods lack interpretability. We apply GIFT to the PanCan12 dataset, and \gift reveals significant relations between cancers, gene sets, and genes, such as influential gene sets for specific cancer (e.g., interferon-gamma response gene set for ovarian cancer) or relations between cancers and genes  (e.g., BRCA cancer $\leftrightarrow$ APOA1 gene and OV, UCEC cancers $\leftrightarrow$ BST2 gene).  \\
\textbf{Availability:} The code and datasets used in the paper are available at \url{https://github.com/leesael/GIFT}. \\
\textbf{Contact:} \href{sael@cs.stonybrook.edu}{sael@cs.stonybrook.edu}\\
\end{abstract}

\maketitle

\footnotetext[1]{These authors contributed equally to this work.}

\section{Introduction}
    \label{sec:intro}
    \begin{figure*}[!htbp]
	\begin{center}
		\includegraphics{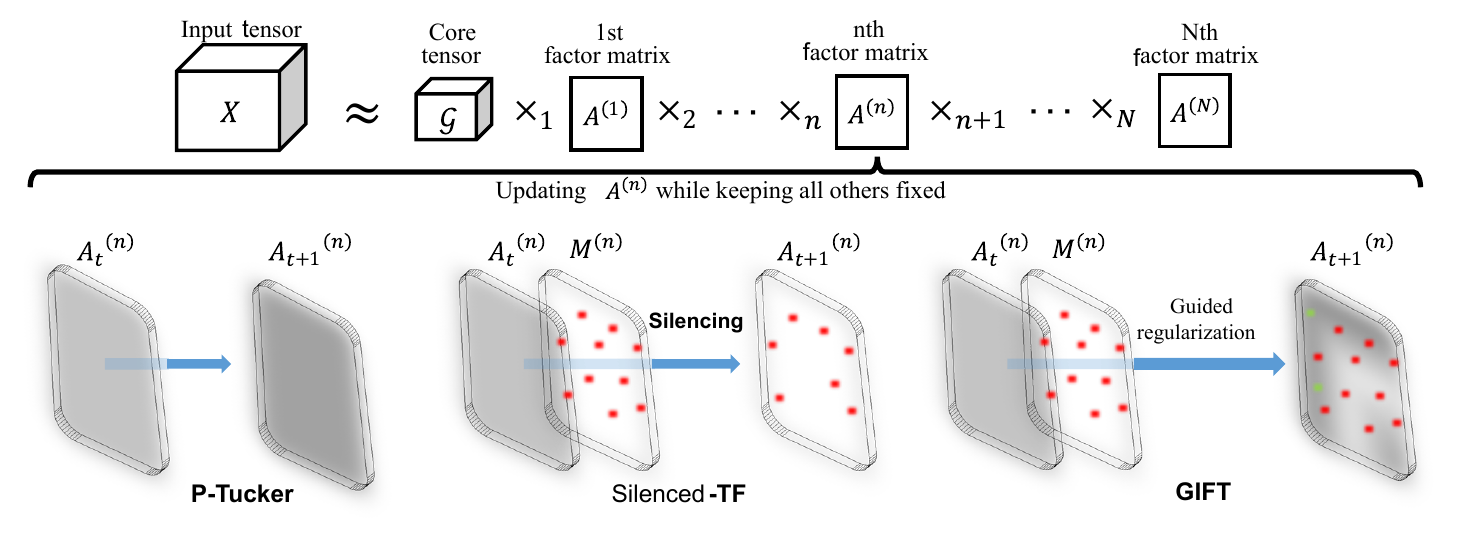}
	\end{center}
	\caption{An overview of \gift, \stf, and \ptf. All methods are based on a Tucker factorization, which decomposes a given tensor into a core tensor and factor matrices. \ptf results in dense gene factor matrices. \stf and \gift result in sparse factor matrices. \stf puts hard constraints while \gift puts soft constraints based on prior knowledge.
	}
	\label{fig:Tucker_factorization}
\end{figure*}

Increasing number of multi-platform genome data of a single person, such as a cancer patient, are being generated.
These data describe different biological aspects of a person and need to be integratively analyzed to obtain a holistic view. However, due to the complexity of the problem, the results of existing methods are difficult to interpret and often do not scale to larger data \citep{Thomas2015}.
Interpretability of the results obtained by the methods is important as most bio-medical studies focus on discoveries.
Scalability of an analysis method is also important as the number of data increases.

\subsection{Integrative Genomic Data Analysis for Cancer Studies}
The Cancer Genome Atlas (TCGA) have reported several integrated genome-wide studies of cancer data. In 2013, TCGA published the PanCan12 dataset that includes multi-platform genomic information of 12 tumor types \citep{weinstein2013cancer}.
The dataset has boosted many genomic cancer analyses \citep{anaya2016pan, riaz2017pan} including the TCGA original multi-platform data analysis \citep{hoadley2014multiplatform}.
\cite{hoadley2014multiplatform} utilizes cluster-of-cluster analysis (COCA) approach for stratification of the Pancan12 dataset.
The COCA is a two-step approach that clusters against cluster results of individual data types.
Although the method is applicable for large data, the two-step process makes it difficult to trace back and interpret the results.
Multi-kernel approaches are also a multi-step approach that the first generates individual kernels from each data type, the second learns a multi-kernel, and the third applies the multi-kernel to cluster or classify \citep{Thomas2017}. Although kernel-based methods are highly accurate, interpretability is lost in the generation of the kernels.
Another integrative method widely used by TCGA is PARADIGM \citep{Vaske2010, Kandoth2013,hoadley2014multiplatform}.
The method is based on Bayesian network inference and is dependent on the biological pathway used and protein expression data. Due to these requirements, it is often applied to a small number of genes.



\subsection{Matrix/Tensor Mining Methods}
Matrix factorization methods, such as the non-negative matrix factorization (NMF), are broadly used across multiple domains, including a bio-data analysis, to analyze data represented as matrices.
NMF was also used extensively by the TCGA group \citep{Koboldt2012, hoadley2014multiplatform, Kandoth2013} and others \citep{Hofree2013, kim2015mutation, Zhu2017} for studying single-platform genome analyses such as somatic mutations or gene expressions.

Natural extensions of single data type modeled as matrices to multi-platform data type are tensors, i.e.,  multi-dimensional arrays.
Tensors are widely applied to represent many real-world data such as movie rating and network traffic data expressed as 3-order tensors with (movie - user - time)  and (source IP - destination IP - time) triples, respectively. Multi-platform genome data can also be represented as a 3-order tensor that contains the experiment values indexed with (patient - gene - experimental platform).

Tensor factorization (TF) methods are applied to analyze tensor data just as matrix factorization methods are used for analyzing matrices.
TF decomposes a given tensor into factor matrices and a core tensor that contains latent relationships between original components.
Applications of TF include anomaly detection from network traffic data \citep{DBLP:conf/kdd/2006}, healthcare monitoring from sensor data \citep{DBLP:journals/corr/WangYM17}, fraud detection from social network data, and biomedical data \citep{cichocki2013tensor, Kim2017}.

However, tensor factorization methods have not been extensively applied to genomic data mainly due to scalability, missing data problem, and interpretability.
For example, the PanCan12 dataset forms a 3-order tensor of size $4,555 \times 14,351 \times 5$. A regular tensor decomposition method will not run due to intermediate data explosion that occurs in the calculation \citep{Jeon2016}.
We have previously addressed the scalability and missing data problems \citep{choi2017fast, Oh2017, ShinK17} and various ways to exploit prior knowledge to obtain high-quality factorizations or intended latent patterns \citep{choi2017fast, Choi2017SNECT, jeon2016scout, Lee2017CTD}. However, factor matrices produced by existing methods are hard to interpret due to their density and unclear value distributions.

Developing an interpretable TF method is essential for analyzing its resultant factors more effectively; poor interpretability makes it hard to discover latent patterns.
The main challenge is to make factor matrices interpretable while preserving the TF accuracy.
Our goal is to devise an interpretable TF method for partially observed tensors exploiting prior knowledge while preserving the accuracy and scalability.

\subsection{Contributions}
\begin{table}[!htbp]
\caption{Comparison of \gift and the other algorithms. \gift produces interpretable results while maintaining high accuracy and scalability. However, \ptf and \stf lack interpretability or accuracy, respectively. \label{tab:comparison}}
{
    \begin{tabular}{c c c c}
                \toprule
                Method & \ptf~\cite{Oh2017} & \stf & \textbf{\gift}  \\
                \midrule
                \textbf{Interpretability}& & \checkmark & \checkmark \\
                \textbf{Accuracy} & \checkmark &  & \checkmark \\
                \textbf{Scalability}& \checkmark & \checkmark & \checkmark \\
                \bottomrule
    \end{tabular}
}
\end{table}

Our main contributions are as follows.
\begin{itemize}
    \item \textbf{Method.} We propose \gift (Guided and  Interpretable Factorization for Tensors) that outputs interpretable factor matrices by constraining the factor matrices based on prior knowledge.
    \item \textbf{Experiments.} We validate that \gift is not only interpretable but highly scalable and accurate (Table~\ref{tab:comparison}).
    \item \textbf{Discovery.} We apply \gift to large-scale multi-platform cancer genome analysis using the PanCan12 dataset and show how the method easily and successfully discovers significant relations between genes, gene sets, and patients, as summarized in Table~\ref{tab:discovery_result}.
\end{itemize}

\section{Methods}
\label{sec:method}
    In this section, we describe datasets used, preliminaries of tensors, and proposed methods.
Preliminaries contain tensor basics and a brief description of our baseline approach \ptf \citep{Oh2017} that we have recently proposed. 
In the proposed methods, we first describe naive interpretable approach \stf and then suggest more advanced method \gift.

\subsection{Data Processing}
We use PanCan12 \citep{weinstein2013cancer} and Hallmark gene set data from MSigDB \citep{Liberzon11MSIGDB} collections as an input tensor and a mask matrix, respectively.
Table \ref{tab:data_table} summarizes the data we used in this paper.

\begin{table}[!htbp]
\caption{Summary of datasets used for experiments. M: million, K: thousand.    \label{tab:data_table}}
{
	\small
    \begin{tabular}{c c c c}
        \toprule
        Dataset & Order & Dimension  & Observable Entries\\
        \midrule
        PANCAN12 tensor & 3 & (4555, 14351, 5) & 180M \\
        Sampled-PANCAN12 & 3 & (4555, 14351, 5) & 36$-$144M \\
        Mask matrix $\mathbf{M}^{(2)}$& 2 & (14351, 50) & 7K \\
        \bottomrule
    \end{tabular}
}
\end{table}

\subsubsection{Mask Matrix}\label{subsubsec:mask}
We generate a gene group mask matrix $\mathbf{M}^{(2)}$ in a form of (gene - gene set) using the Hallmark data, which contain 50 important gene sets.
Each column of mask matrix $\mathbf{M}^{(2)}$ corresponds to a gene set.
If a gene $i_n$ is contained in a gene set $j_n$ then it is unmasked, i.e., $\mathbf{M}_{i_n j_n}^{(2)}$ is set to $0$; otherwise, the gene is masked, i.e., set to 1.
If no prior-knowledge is known, the mask matrices are set to zero matrices with the size of corresponding factor matrices.

\subsubsection{PanCan12 Tensor}\label{subsubsec:pancan12}
The PanCan12 dataset is represented as a 3-order tensor in a form of (patient - gene - platform; experiment value).
Initially, the 4.7 version of the PanCan12 was downloaded from the Sage Bionetworks repository, Synapse \citep{Omberg2013}.
The PanCan12 contains multi-platform data with mapped clinical information of patients group into cohorts of twelve cancer type: bladder urothelial carcinoma (BLCA), breast adenocarcinoma (BRCA), colon and rectal carcinoma (COAD, READ), glioblastoma multiforme (GBM), head and neck squamous cell carcinoma (HNSC), kidney renal clear cell carcinoma (KIRC), acute myeloid leukemia (LAML), lung adenocarcinoma (LUAD), lung squamous cell carcinoma (LUSC), ovarian serous carcinoma (OV), and uterine corpus endometrial carcinoma (UCEC).
After download, probes of each platform are mapped to corresponding gene symbols.
Then, subjects that have less than two evidence are removed from the dataset and genes that are not part of Hallmark set are removed.
The resulting data for each platform are min-max normalized and is further normalized such that the Frobenius norm, i.e.,$\|A\|\equiv\sqrt{\sum_i\sum_j|a_{ij}|^2}$.

\subsection{Tensor Preliminaries}
\label{secS:prelim}
In this section, we describe preliminaries of a tensor and its factorization methods. Table \ref{tab:symbol_table} summarizes symbols used in this paper.

\begin{table}[!htp]
\caption{Table of symbols.     \label{tab:symbol_table}}
{
	\small
    \begin{tabular}{cl}
        \toprule
        \textbf{Symbol} & \textbf{Definition} \\
        \midrule
        $\T{X}$ & tensor (Euler script, bold letter) \\
        $\mathbf{X}$ & matrix (uppercase, bold letter) \\
        $x$ & scalar (lower case, italic letter)\\
        $N$ & order (number of modes) of a tensor\\
        $I_{n},J_n$ & dimensionality of the $n$th mode of input and core tensor\\
        $\mat{A}^{(n)}$ & $n$th factor matrix $(\in \mathbb{R}^{I_{n} \times J_{n}})$ \\
        $a^{(n)}_{i_{n}j_{n}}$ & $(i_{n},j_{n})$th entry of $\mat{A}^{(n)}$\\
        $\Omega$ & set of observable entries of $\T{X}$\\
        $|\Omega|,|\T{G}|$ & number of observable entries of input and core tensor\\
        $\lambda$ & regularization parameter for factor matrices\\
        $||\bullet||_F$ & Frobenius norm \\
        $\ast$ & element-wise multiplication\\
        $\circ$ & outer product \\
        ${\times}_{n}$ & n-mode product \\
        \bottomrule
    \end{tabular}
}{}
\end{table}

\subsubsection{Tensor}
A tensor is a multi-dimensional array which is a generalization of a matrix and a vector.
A mode or a way indicates each axis of a tensor, and an order is the number of modes or ways.
We denote a tensor using boldface Euler script letters (e.g., $\T{X}$).
A tensor $\T{X} \in \mathbb{R}^{I_1 \times I_2 \times \dots \times I_N}$ is an $N$-order tensor which has $N$ modes whose lengths are from $I_1$ to $I_N$.
A vector and a matrix are regarded as a $1$- and a $2$-order tensor, respectively.
We denote a matrix and a vector using boldface uppercase (e.g., $\mathbf{X}$) and lowercase letters (e.g., $\mathbf{x}$), respectively.
The $i_{1}$th row of $\mat{A}$ is denoted by $\vect{a}_{i_{1}:}$, and the $i_{2}$th column of $\mat{A}$ is denoted by $\vect{a}_{:i_{2}}$.

\subsubsection{Tensor Decomposition}
Among many tensor decomposition methods, we use Tucker factorization~\cite{tucker1966some, de2000multilinear} methods, which allows us to discover not only latent concepts but also relations between the concepts hidden in tensors~\cite{Oh2017}.
Tucker factorization decomposes a given tensor $\T{X}$ into a core tensor $\T{G}$ and factor matrices $\mathbf{A}^{(1)}, \cdots , \mathbf{A}^{(N)}$, as defined in Definition \ref{def:Tucker_factorization}.
\begin{definition} \label{def:Tucker_factorization}
    \textbf{(Tucker factorization)}
    Given a tensor $\T{X} \in \mathbb{R}^{I_1 \times I_2 \times \dots \times I_N}$, the Tucker factorization of rank $(J_1, \cdots , J_N)$ finds a core tensor $\T{G} \in \mathbb{R}^{J_1 \times \cdots \times J_N}$ and factor matrices $\mathbf{A}^{(1)} \in \mathbb{R}^{I_1 \times J_1}, \cdots ,\mathbf{A}^{(N)} \in \mathbb{R}^{I_N \times J_N}$, which minimize the following objective function~\eqref{eq:Tucker}.
    \begin{equation} \label{eq:Tucker}
    \small
    \hspace{-2mm}
    \begin{aligned}
    &\mathcal{L}(\T{G}, \mathbf{A}^{(1)}, \cdots ,\mathbf{A}^{(N)}) \\
    &= || \T{X} - \sum_{\forall(j_1,...,j_N) \in \T{G}} \T{G}_{j_1, ..., j_N} (\mathbf{a}_{j_1}^{(1)} \circ \cdots \circ \mathbf{a}_{j_N}^{(N)})||_F^2 \\
    &= \sum_{\forall(i_1, ..., i_N) \in \T{X}} \Bigg( \T{X}_{(i_1, ... , i_N)} - \sum_{\forall(j_1,...,j_N) \in \T{G}} \T{G}_{(j_1,..., j_N)} \prod_{n=1}^{N} a_{i_n j_n}^{(n)} \Bigg)^2
    \end{aligned}
    \end{equation}
\end{definition}
Note that Equation~\eqref{eq:Tucker} assumes missing entries of $\T{X}$ as zeros.
Each column vector of a factor matrix generally represents each different concept.
A higher value in a vector indicates that the corresponding element is highly related to the concept.
Assuming a given tensor is movie rating data with (movie - user - time) triples, then a column vector in a movie-factor matrix can have a concept such as a horror or comic genre.

\subsubsection{Partially Observable Tensor Factorization}
Many real-world tensors have missing values in them (i.e. partially observed).
Applying standard Tucker factorization methods to the data triggers highly inaccurate results since they regard missing entries as zeros.
Partially observable tensor factorization methods focus only on observed entries to tackle this problem, and a partially observable Tucker factorization is defined as follows.

\begin{definition}\label{TF_for_partially_observed_tensor}
    \textbf{(Partially Observable Tucker Factorization)}
    Given a tensor $\T{X} \in \mathbb{R}^{I_1 \times I_2 \times \dots \times I_N}$ with observable entries $\Omega$, a partially observable Tucker factorization of rank $(J_1, \cdots , J_N)$ finds a core tensor $\T{G} \in \mathbb{R}^{J_1 \times \cdots \times J_N}$ and factor matrices $\mathbf{A}^{(1)} \in \mathbb{R}^{I_1 \times J_1}, \cdots ,\mathbf{A}^{(N)} \in \mathbb{R}^{I_N \times J_N}$ which minimize the following objective function.
    \begin{equation} \label{eq:obj_func_TF_for_partially_observed}
    \begin{aligned}
    &\mathcal{L}(\T{G}, \mathbf{A}^{(1)}, \cdots ,\mathbf{A}^{(N)}) = \\
    &\sum_{\forall(i_1, ..., i_N) \in \T{X}} \Bigg( \T{X}_{(i_1, ... , i_N)} - \sum_{\forall(j_1,...,j_N) \in \T{G}} \T{G}_{(j_1,..., j_N)} \prod_{n=1}^{N} a_{i_n j_n}^{(n)} \Bigg)^2 \\
    &+ \lambda \Bigg(\sum_{n = 1}^{N}  ||\mathbf{A}^{(n)}||_{F}^2 \Bigg)
    \end{aligned}
    \end{equation}
    Note that $\lambda$ denotes a regularization parameter for factor matrices, and we used $L_2$-regularization to prevent overfitting, which has been widely used in recommender systems~\cite{koren2009matrix, ShinK17}.
\end{definition}

\subsubsection{Baseline Approach: \ptf} 
\label{sec:met:ptf}
Among many Tucker factorization methods~\cite{smith2017tucker, Oh:2017:SHOT, filipovic2015tucker}, \ptf ~\cite{Oh2017} shows the best scalability and accuracy for partially observable tensors by focusing on observed entries of the tensors.
The objective function of \ptf is the same to Equation~\eqref{eq:obj_func_TF_for_partially_observed}, and \ptf uses a row-wise alternating least squares (ALS) to minimize the loss function.
In detail, \ptf first chooses a row of a factor matrix to be updated while fixing all the others, and it computes three intermediate data $\delta_{(i_1,...,i_N)}^{(n)}$, $\mathbf{B}_{i_n}^{(n)}$, and $\vect{c}_{i_n:}^{(n)}$ defined as follows. Notice that $\mat{I}_{J_n}$ is a $J_n \times J_n$ identity matrix.

$\delta_{(i_1,...,i_N)}^{(n)}$ is a length ${J_n}$ vector whose $j$th entry is
\begin{equation} \label{eq:delta}
\sum_{\forall(j_1...j_n=j...j_N)\in\T{G}}\T{G}_{(j_1...j_n=j...j_N)}\prod_{k \neq n} a^{(k)}_{i_k j_k},
\end{equation}
$\mat{B}_{i_n}^{(n)}$ is a  ${J_n \times J_n}$ matrix whose $(j_1,j_2)$th entry is
\begin{equation} \label{eq:rowB}
\sum_{\forall(i_1,...,i_N)\in\Omega_{i_n}^{(n)}}\delta_{(i_1,...,i_N)}^{(n)}(j_1) \delta_{(i_1,...,i_N)}^{(n)}(j_2),
\end{equation}
and $\vect{c}_{i_n:}^{(n)}$ is a length ${J_n}$ vector whose $j$th entry is
\begin{equation} \label{eq:rowC}
\sum_{\forall(i_1,...,i_N)\in\Omega_{i_n}^{(n)}}\T{X}_{(i_1,...,i_N)}  \delta_{(i_1,...,i_N)}^{(n)}(j).
\end{equation}

Using the above intermediate data, \ptf updates a row $a_{i_n :}^{(n)}$ by an update rule $\vect{c}_{i_n:}^{(n)} \times [\mat{B}_{i_n}^{(n)}+\lambda \mathbf{I}_{J_n}]^{-1}$. After updating factor matrices, \ptf calculates reconstruction error by the following rule.

\begin{equation} \label{eq:recon}
\small
\sqrt{\sum_{\forall(i_1, ..., i_N) \in \Omega} \Bigg( \T{X}_{(i_1, ... , i_N)} - \sum_{\forall(j_1,...,j_N) \in \T{G}} \T{G}_{(j_1,..., j_N)} \prod_{n=1}^{N} a_{i_n j_n}^{(n)} \Bigg)^2}
\end{equation}

If the error converges or the maximum iteration is reached, \ptf stops iterations and performs QR decompositions to orthogonalize factor matrices and update a core tensor accordingly. Note that \cite{Oh2017} suggests full details and proofs of the update process.

\subsection{Proposed Methods}
The main challenge is to devise a method which employs prior knowledge to produce interpretable factors while maintaining the decomposition quality and speed.


\subsubsection{\stf}
Although our previous method \ptf presents high scalability and accuracy, it is hard to interpret the results of \ptf since the factor matrices are dense and there is no direct interpretable connection between significant components, i.e. genes in each factor column that describes a group.
\stf literally silences uninteresting or unnecessary parts of factor matrices and updates the rest using the same algorithm of \ptf.

More specifically, given a tensor $\T{X} \in \mathbb{R}^{I_1 \times I_2 \times \dots \times I_N}$ with observable entries $\Omega$, \stf of rank $(J_1, \cdots , J_N)$ finds a core tensor $\T{G} \in \mathbb{R}^{J_1 \times \cdots \times J_N}$ and factor matrices $\mathbf{A}^{(1)} \in \mathbb{R}^{I_1 \times J_1}, \cdots ,\mathbf{A}^{(N)} \in \mathbb{R}^{I_N \times J_N}$ which minimize the following objective function subjected to mask matrices $\mathbf{M}^{(1)} \in \mathbb{R}^{I_1 \times J_1}, \cdots ,\mathbf{M}^{(N)} \in \mathbb{R}^{I_N \times J_N}$.


\begin{equation}
\begin{aligned}
& \underset{\T{G}, \mathbf{A}^{(1)}, \cdots ,\mathbf{A}^{(N)}}{\text{minimize}}
& & \mathcal{L}(\T{G}, \mathbf{A}^{(1)}, \cdots ,\mathbf{A}^{(N)},\mathbf{M}^{(1)}, \cdots ,\mathbf{M}^{(N)}) 
\\
& \text{subject to}
& & a_{i_n j_n}^{(n)} = 0 \; when \;\; m_{i_n j_n}^{(n)} = 1
\end{aligned}
\end{equation}

The difference in \stf compared to \ptf is in the selective updates of rows of factor matrices.
To be more specific, given the mask matrices that encode the prior knowledge, \stf only updates an entry in the gene factor matrix, $a_{i_n j_n}^{(n)}$, when the corresponding masking element $m_{i_n j_n}^{(n)}$ is 0, and \stf sets the entry  $a_{i_n j_n}^{(n)}$  in the gene factor matrix as a zero when $m_{i_n j_n}^{(n)}$ is 1.
By masking, \stf offers interpretable gene factor matrices in addition to the benefits offered by \ptf: scalability and applicability for partially observed data.

\subsubsection{\gift}
Two potential weaknesses of \stf are low factorization accuracy due to many zeros in its factor matrices and inability to find genes that have a significant relationship to the gene group if they are specified as a member the gene group.
To address these weaknesses, we propose \gift which offers interpretability, high accuracy, and flexibility.
\gift tackles the problem by employing selective regularization of factor matrices. \gift penalties proportional to $\lambda$ on masked entries of factor matrices during the update process. Thus, \gift makes a distinction between the values of masked and unmasked genes for each gene group. Moreover, the accuracy of \gift is similar to \ptf since masked entries can also have small values unlike the strict zero constraints of \stf.

\begin{algorithm} [t!]
    \small
    \caption{GIFT} \label{alg:GIFT}
    \begin{algorithmic}[1]
        \REQUIRE
        A tensor $\T{X} \in \mathbb{R}^{\mathit{I}_1 \times \cdots \times \mathit{I}_N}$ with observable entries $\Omega$, mask matrices $\mathbf{M}^{(1)}, \cdots ,\mathbf{M}^{(N)}$, rank $(J_1, \cdots , J_N)$, and a regularization parameter $\lambda$.
        
        \ENSURE A core tensor $\T{G}$ and factor matrices $\mathbf{A}^{(1)}, \cdots \mathbf{A}^{(N)}$.
        
        \STATE initialize $\T{G}$ and $\mathbf{A}^{(1)}, \cdots, \mathbf{A}^{(N)}$ randomly
        
        \REPEAT
        
        \FOR{$n = 1, \cdots , N$}
        
        \FOR{$i_n = 1, \cdots , I_n$}
        
        \STATE calculate intermediate data  $\delta$, $\mathbf{B}_{i_n}^{(n)}$, and $\vect{c}_{i_n:}^{(n)}$ by \eqref{eq:delta} -- \eqref{eq:rowC}
        
        \STATE calculate $\mathbf{D}_{i_n}^{(n)}$, where its $(j_n,j_n)$th entry is $\mathbf{M}_{i_n j_n}^{(n)}$
        
        \STATE update a row $a_{i_n :}^{(n)}$ by $\vect{c}_{i_n:}^{(n)} \times [\mat{B}_{i_n}^{(n)}+\lambda \mathbf{D}_{i_n}^{(n)}]^{-1}$
        
        \ENDFOR
        
        \ENDFOR
        
        \STATE {compute reconstruction error by \eqref{eq:recon}}
        
        \UNTIL{a convergence criterion is met}        
        
        \FOR{$n = 1 ... N$}
        \STATE $\mathbf{A}^{(n)} \rightarrow \mathbf{Q}^{(n)}\mathbf{R}^{(n)}$\\
        \STATE $\mathbf{A}^{(n)} \leftarrow \mathbf{Q}^{(n)}$\\
        \STATE $\T{G} \leftarrow \T{G}\times_{n}\mathbf{R}^{(n)}$
        
        \ENDFOR
    \end{algorithmic}
\end{algorithm}

\gift encodes mask matrices $\mathbf{M}^{(n)}$ into its objective function as a regularization term.
The specific loss function of GIFT is given by the following Equation~\eqref{eq:obj_func_of_GIFT}.
\begin{equation}
\label{eq:obj_func_of_GIFT}
\begin{aligned}
&\mathcal{L}(\T{G}, \mathbf{A}^{(1)}, \cdots ,\mathbf{A}^{(N)},\mathbf{M}^{(1)}, \cdots ,\mathbf{M}^{(N)}) = \\
&\sum_{\forall(i_1,...,i_N) \in \Omega} \Bigg( \T{X}_{(i_1,\cdots,i_N)} - \sum_{\forall(j_1,...,j_N) \in \T{G}}  \T{G}_{(j_1,\cdots,j_N)} \prod_{n=1}^{N} a_{i_n j_n}^{(n)} \Bigg)^2 \\
&+ \lambda \Bigg( \sum_{n = 1}^{N}  ||\mathbf{M}^{(n)} \ast \mathbf{A}^{(n)}||^2 \Bigg)
\end{aligned}
\end{equation}
The main difference between \ptf and \gift is an existence of mask matrices $\mathbf{M}^{(n)}$ in a regularization term.
GIFT uses $\mathbf{M}^{(n)} \ast \mathbf{A}^{(n)}$ instead of just $\mathbf{A}^{(n)}$, where $\ast$ denotes an element-wise multiplication.
Through the specially-designed regularization, \gift fully exploits prior knowledge encoded in $\mathbf{M}^{(n)}$.
Compared to \stf, \gift shows flexibility regarding the updates of masked entries. Instead of fixing them as zeros, \gift imposes regularizations on them, which tend to make the values smaller, but not normally zeros.

Algorithm \ref{alg:GIFT} describes how \gift updates given factor matrices. When \gift updates a row $a_{i_n :}^{(n)}$ (line 6), it requires a diagonal matrix $D$ which reflects masking information (line 5), while \ptf uses an identity matrix $\mat{I}_{J_n}$. The other parts of \gift are the similar to that of \ptf.
Table \ref{tab:comparison} shows a comparison of \gift, \stf, and \ptf.

\section{Results}
    \label{sec:experiment}
    In this section, we describe experimental results of \gift compared to \stf and \ptf.
We aim to answer the following questions.

\noindent \textbf{[Q1] Interpretability}: How interpretable are factor matrices produced by \gift and the other methods? (Section \ref{sec:exp:interpretability})\\
\noindent \textbf{[Q2] Accuracy}: How accurately do \gift and the other methods factorize a given tensor and predict missing entries of the tensor?  (Section \ref{sec:exp:accuracy}) \\
\noindent \textbf{[Q3] Scalability}: How well do \gift and the other methods scale up with respect to the number of observed entries of a tensor? (Section \ref{sec:exp:scalability})

\subsection{Experimental Settings}

\gift , \stf, and \ptf are implemented in C with \textsc{OpenMP} and \textsc{Eigen} libraries. We run our experiments
on a single machine with 20 cores / 40 threads, equipped with
an Intel Xeon E5-2630 v4 2.2GHz CPU and 512GB RAM.
We set the default rank as $(30\times50\times2)$.
In reporting running times, we use the average elapsed time per iteration, not the total running time.
Notice that we use absolute values of factor matrices for all experiments.

\subsection{Interpretability} \label{sec:exp:interpretability}
\begin{figure}[!htbp]
\includegraphics[width=0.45 \textwidth]{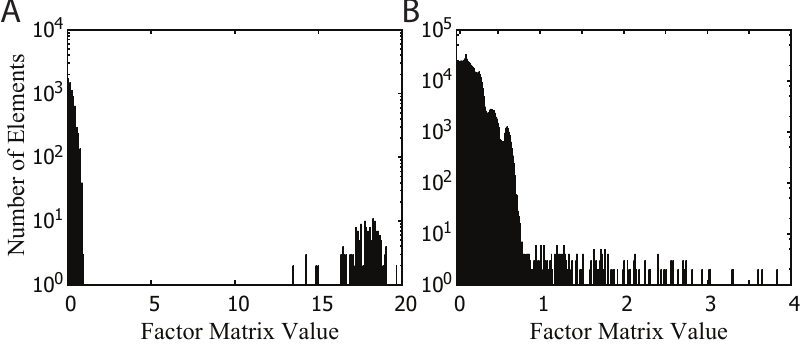}
    \caption{Distributions of values in a gene factor matrix derived by \gift ($\lambda$ = 10) for unmasked (\textbf{A}) and masked (\textbf{B}) entries.  }
    \label{fig:exp:distribution}
\end{figure}

We regard a gene factor matrix as interpretable if the genes composing a gene set (unmasked) have distinguishably larger factor values than the factor values of genes that are not in the gene set (masked).
In \stf, this is strictly enforced by making the factor values of genes that are not in the gene set (masked) are all set to zeros.
In \gift, this is not strictly enforced. In other words, there is a potential that factor values of the masked genes become large and those of the unmasked genes become small. This is likely to happen only when the prior knowledge and the data do not agree.
However, as presented in Figure~\ref{fig:exp:distribution}, \gift generates a gene factor matrix within high factor values (larger than 10) for the subset of unmasked entries (\textbf{A}) and small factor values (less than 1) for a majority of the masked entries (\textbf{B}). This shows that \gift has learned the latent relationships of cancer patients to gene sets and significant genes in the gene set.
\ptf, on the other hand, produces a gene factor matrix with value distribution that has small or no correlation to a gene set (refer to the supplementary material for the results of \stf and \ptf).

\begin{figure}[!htbp]
    \centering
    \includegraphics{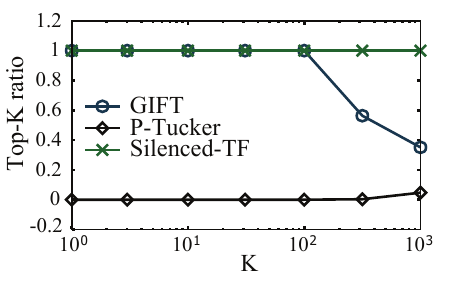}
    \caption{Top-K ratios of \gift and the other methods. A high top-K ratio implies that the corresponding method is interpretable.}
    \label{fig:exp:topk}
\end{figure}

Additionally,  top-K ratios can be used as a metric of interpretability.
A top-K ratio indicates how many unmasked entries are included in total top-K entries (in descending order by values) of a gene factor matrix, which is defined as follows.
\begin{equation}
\text{Top-K ratio R }(0 \leq R \leq 1) = \frac{\text{number of unmasked entries in top-K}}{K}
\end{equation}
Figure~\ref{fig:exp:topk} illustrates top-K ratios.
\ptf shows the worst top-K ratios for all K since it does not distinguish unmasked and masked entries in the calculation.
Although \stf exhibits the highest top-K ratios for all K by silencing the masked entries, \stf cannot discover important masked entries which are closely related to unmasked entries since their values are all set to zeros.
Meanwhile, the top-K ratio of \gift is the highest until $K \leq 10^2$ and decreases rapidly after $K \geq 10^2$ when the important unmasked entries are saturated and top values start including the unimportant unmasked entries and important masked entries.
Overall, \stf and \gift provide interpretable factorizations with respect to distributions of values in a factor matrix and top-K ratios.

\begin{figure*}[!ht]
\includegraphics[width=0.9\textwidth]{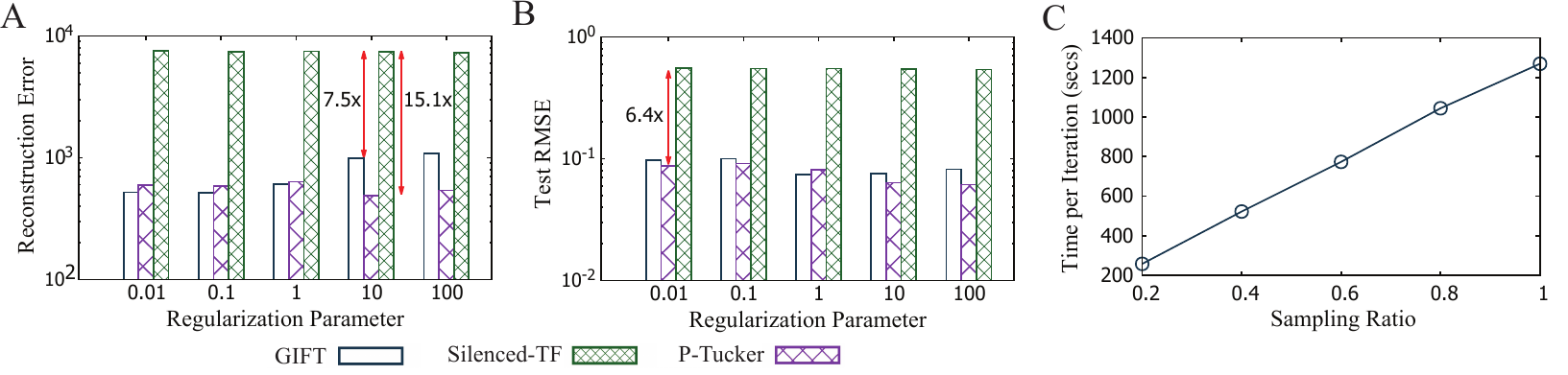}
\centering
\caption{Performance comparisons of \gift, \stf, and \ptf.
\textbf{A} is a reconstruction error plot. \stf is highly inaccurate due to many zeros in its factor matrix, while \gift and \ptf present a high accuracy.
\textbf{B} is a test RMSE plot. \stf is highly inaccurate due to many zeros in its factor matrix, while \gift and \ptf present a high accuracy.
\textbf{C} is a scalability plot of \gift with respect to the number of observable entries in the tensor. As the number of observed entries increases, a running time of \gift increases proportionally.}
\label{fig:exp:performance}
\end{figure*}

\subsection{Accuracy} \label{sec:exp:accuracy}
We use two evaluation metrics---reconstruction error and test root means square error (RMSE)---to measure the accuracy. 
Reconstruction error indicates an accuracy of a factorization as given in Equation~\eqref{eq:recon} and test RMSE implies how accurately a method predicts missing entries of a tensor. To measure test RMSE, we split the PANCAN12 tensor into training/test data with a ratio of 9 to 1.
As illustrated in Fig.~\ref{fig:exp:performance}\textbf{A} and Fig.~\ref{fig:exp:performance}\textbf{B}, \stf exhibits the worst accuracy due to many zeros in a silenced factor matrix. The reconstruction error and test RMSE of \stf are $15.1 \times$ and $6.4 \times$ higher than that of \ptf when $\lambda=10$ and $\lambda=0.01$, respectively. While \ptf shows the best accuracy in most cases, \gift presents relatively small accuracy loss compared to that of \stf; in particular, test RMSE of \gift is slightly higher or even better than that of \ptf.

\subsection{Scalability} \label{sec:exp:scalability}
Scalability test is performed by varying the number of observable entries by randomly sampling 20\%, 40\%, 60\%, 80\%, and 100\% from the PANCAN12 tensor. 
As shown in  Fig.~\ref{fig:exp:performance}\textbf{C}, \gift scales near linearly in terms of the number of observable entries. We omit results of \ptf and \stf since they present similar scalability to that of \gift (refer to the supplementary material). Note that computational time and memory usage increase exponentially in regular tensor decompositions. 

\subsection{Discovery}
\label{sec:discovery}
In this section, we describe our finding on (cancer - gene sets), (gene sets - genes), and (cancer - genes) relations hidden in the PANCAN12 dataset by interpreting results of \gift with $\lambda=10$.

\begin{table*}[!htbp]
	\caption{Discoveries on the PANCAN12 dataset via \gift. \gift extracts significant gene sets and notable relations between cancer, gene sets, and genes.
		Many biological research results substantiate the retrieved relations, which are described in a gene description column. ( $^*$: important gene, $^-$: unimportant gene, $^+$: not included in a gene set, but related).
		\label{tab:discovery_result}}
	{
		\begin{tabular}[t]{C{1.5cm} C{2cm} C{1.5cm} C{1.5cm} C{9.5cm}}
			\hline
			\textbf{Cancer} & \textbf{Gene set} & \textbf{Genes} & \textbf{Factor} & \textbf{Gene Description} \\
			
			\hline
			\multirow{3}{*}{\shortstack{HNSC, LUAD, \\LUSC, BLCA}} & \multirow{3}{*}{\shortstack{TGF beta signaling}} & SKIL$^*$ & 0.5146 & Encodes the SNON, negative regulators of TGF-beta signaling \cite{tecalco2012transforming}. \\
			& & FKBP1A$^*$ & 0.4692 & Interacts with a type I TGF-beta receptor. \\
			& & LEFTY2$^*$ & 0.2925 & Encodes a secreted ligand of the TGF-beta family of proteins. \\
			
			\hline
			\multirow{3}{*}{\shortstack{GBM}} & \multirow{3}{*}{\shortstack{Angiogenesis}} & PF4$^*$ & 0.5049 & Inhibits cell proliferation and angiogenesis in vitro and in vivo \cite{bikfalvi2004platelet}. \\
			& & VCAN$^*$ & 0.4500 &  Encodes a protein involving in cell adhesion, and angiogenesis \cite{wight2002versican}. \\
			& & LPL$^-$ & 0.0429 & Encodes lipoprotein lipase \cite{wang2009lipoprotein}.\\
			
			\hline
			\multirow{5}{*}{\shortstack{BRCA}} & \multirow{3}{*}{\shortstack{Estrogen response\\ late}} & IL17RB$^*$ & 0.3807 & Involved in development and progression of breast cancer \cite{alinejad2017}. \\
			& & TFF3$^*$ & 0.3640 & Promotes invasion and migration of breast cancer \cite{may2015tff3}. \\
			& & PTGER3$^-$ & 0.0200 & Encode protein related to digestion and nervous system. \\ \cline{2-5}
			& \multirow{2}{*}{\shortstack{Bile acid\\ metabolism}} & \multirow{2}{*}{\shortstack{APOA1$^*$}} & \multirow{2}{*}{\shortstack{0.3973}} & \multirow{2}{*}{\shortstack{Encodes lipoprotein lipase, an enzyme which hydrolyzes lipoprotein. \\Known breast cancer risk factor \cite{martin2015serum}. }}\\
			& & & & \\
			\hline
			
			\multirow{4}{*}{\shortstack{OV, UCEC}} & \multirow{3}{*}{\shortstack{Interferon-gamma\\ response}} & IRF7$^*$ & 0.5727 &Encodes interferon regulatory factor 7. \\
			& & BST2$^*$ & 0.4986 & High levels of BST2 have been identified in ovarian cancer \cite{shigematsu2017overexpression}. \\
			& & SSPN$^-$ & 0.0983 & Associated with a skeletal muscle membrane \cite{lapidos2004dystrophin}. \\ \cline{2-5}
			
			& Apoptosis & CASP8AP2$^+$ & 1.4708 & Associated with apoptosis of leukemic lymphoblasts \cite{flotho2006genes}.\\
			&&&& Encoded protein plays a regulatory role in Fas-mediated apoptosis ~\cite{pmid10235259}. \\
			
			\hline
			READ, COAD & Protein secretion & STX7$^*$ & 0.4854 & Controls vesicle trafficking events involved in cytokine secretion \cite{achuthan2008regulation}.\\
			
			\hline
			KIRC, LAML & Mitotic spindle & LATS1$^*$ & 0.4913 & Binds phosphorylated zyxin and moves it to the mitotic spindle. \\
			\bottomrule
		\end{tabular}
	}{}
	
\end{table*}

\subsubsection{Cancer to gene sets findings}
Given specific cancer type, which gene set is the most relevant to the cancer type?
We first explain our discovery procedure and introduce several examples of (cancer - gene sets) relations found by \gift.

We first compute an influence of each gene set on a patient and extract top-$k$ important gene sets for the patient. After that, we aggregate all top-$k$ gene sets of patients suffering from the given cancer and derive top-$k$ relevant gene sets to cancer by choosing top-$k$ frequent gene sets in aggregations. In detail,
$\mathbf{a}_{i:}^{(1)}$ is a latent feature of $i$-th patient, and $\mathbf{G} = (\sum_{i = 1}^{I_3}\T{G}_{::i}) / I_3$ is a relation between gene sets and  columns of a patient-factor matrix.
Then, we use $\tilde{\mathbf{a}}_{i:}^{(1)} = \mathbf{a}_{i:}^{(1)} \mathbf{G}$ as an influence of each gene set on the $i$-th patient.
The $j$-th element of $\tilde{\mathbf{a}}_{i:}^{(1)}$ indicates the influence of $j$-th gene set on the $i$-th patient.
We extract top-$k$ most important gene sets for each patient by selecting top-$k$ highest values in $\tilde{\mathbf{a}}_{i:}^{(1)}$.
Finally, we count the frequency of gene sets that appeared in the top-$k$ gene sets of all patients of a given cancer type.
We regard the most frequent gene set as the most relevant one to the given cancer.

The first and second columns of Table \ref{tab:discovery_result} show (cancer - gene sets) relations discovered by \gift.
For breast cancer (BRCA), GIFT considers `Estrogen response late' and `Bile acid metabolism' gene sets closely related to breast cancer.
It is well known that the estrogen plays a key role in the occurrence of breast cancer 
while the relation to `Bile acid metabolism' gene set seems unnatural.
However, \cite{murray1980faecal} reveal that patients with breast cancer have significantly low fecal bile acid concentration than that of controlled patients.
For ovarian cancer (OV), a relation to the `Interferon-gamma response' gene set is supported by \cite{wall2003}.
They show that interferon-gamma causes apoptosis in human epithelial ovarian cancer.
The `TGF beta signaling' gene set is frequent among many types of cancer including Head and Neck Squamous Cell Carcinoma (HNSC), Lung adenocarcinoma (LUAD), Lung Squamous, Cell Carcinoma (LUSC), and Bladder carcinoma (BLCA).
The reason is that the Transforming growth factor-$\beta$ (TGF-$\beta$) gene set is a tumor suppressor which affects many types of human cancers \citep{kretzschmar2000transforming}.
%

\subsubsection{Gene sets to genes findings}
Given a gene set, which genes are important or unimportant within it?
Is there a gene not included but related to the gene set?
A high value in the gene factor matrix indicates that the corresponding gene is highly related to the corresponding gene set.
We sort the genes in each column of the gene factor matrix in descending order by their value and inspect high-value genes for each gene set.

We show how the discovered (gene sets - genes) relations are supported by biological facts with examples.
The second and third columns of Table \ref{tab:discovery_result} show (gene sets - genes) relations retrieved by \gift.
The SKIL gene in the `TGF beta signaling' gene set is identified to be important by \gift.
The SKIL gene encodes a protein which antagonizes TGF-$\beta$ signaling \citep{tecalco2012transforming}.
The PF4 gene in the `Angiogenesis' gene set, reported to be important by \gift, is known as an inhibitor of cell proliferation and angiogenesis \citep{bikfalvi2004platelet}.
The IRF7 gene in the `Interferon-gamma response' gene set is also identified to be important, and the gene encodes interferon regulatory factor 7.
The LPL gene in the `Angiogenesis' gene set is unimportant according to the discovery result of \gift.
The LPL gene encodes lipoprotein lipase, an enzyme which hydrolyzes lipoprotein \citep{wang2009lipoprotein}, thus it has low relatedness to angiogenesis.
\gift also reports the CASP8AP2 gene is closely related to the `Apoptosis' gene set although the gene is not included in the gene set.
The CASP8AP2 gene is associated with apoptosis of leukemic lymphoblasts in reality \cite{flotho2006genes}.

\subsubsection{Cancer to genes findings}
Given specific cancer type, which genes affect the cancer type most?
We suggest (cancer - genes) relations by combining two relations (cancer - gene sets) and (gene sets - genes) discovered by \gift.

The first and third columns of Table \ref{tab:discovery_result} show (cancer - genes) relations found by \gift.
We regard gene sets in the second column of the table as bridges for (cancer - genes) relations.
We deduce the IL17RB and TFF3 genes are significant to breast cancer since the genes are both important for the `Estrogen response late' gene set and the gene set is the most relevant one to breast cancer.
\cite{alinejad2017} showed that IL17RB is crucial in development and progression of breast cancer in effect.
Moreover, \cite{may2015tff3} reveal that the TFF3 gene promotes invasion and migration of breast cancer.
\gift also finds that the APOA1 gene in the `Bile acid metabolism' gene set is highly related to breast cancer.
High levels of APOA1 are known to be related to increased breast cancer risk \citep{martin2015serum}.
In the case of ovarian cancer, \gift asserts a strong relation to the BST2 gene.
High levels of BST2 have been identified in ovarian cancer \citep{shigematsu2017overexpression}.

\section{Discussions and Conclusion}
    \label{sec:conclusion}
    In this paper, we proposed \gift, a guided and interpretable factorization method for tensors.
\gift provides interpretable factor matrices by encoding prior knowledge through selective regularization.
Experiment results demonstrate that \gift produces interpretable factorizations with high scalability and accuracy, while other methods lack accuracy or interpretability.
In practice, we apply \gift to human cancer analytic using the PANCAN12 dataset and successfully identify important relations between cancers, gene sets, and genes.
For instance, \gift suggests influential gene sets for specific cancer (e.g., interferon-gamma response gene set for ovarian cancer).
In addition, \gift discovers relations between cancers and genes  (e.g., BRCA cancer $\leftrightarrow$ APOA1 gene and OV, UCEC cancers $\leftrightarrow$ BST2 gene).
Furthermore, \gift is able to extract out-of-the-box relations, which are not given in prior information.
Specifically, in Hallmark gene set data, a CASP8AP2 gene was not included in a gene set about apoptosis. However, \gift elicits a relation between the gene and gene set, which is an acknowledged relation by \cite{flotho2006genes, pmid10235259}.

\section*{Funding}
This research was supported by Basic Science Research Program through the National Research Foundation of Korea \\ (NRF-2015R1C1A2A01055739) and by the Korea Evaluation Institute Of Industrial Technology under the ``Global Advanced Technology Center'' (10053204).

\bibliographystyle{ACM-Reference-Format}

\clearpage
\newpage

\setcounter{figure}{0}
\setcounter{table}{0}
\setcounter{section}{1}

\section*{Supplementary Information}
\label{sec:appendix}

\subsection{Interpretability} \label{secS:exp:interpretability}

In this section, we offer additional experimental results of \gift and other methods. In detail, we first introduce several value distributions of masked and unmasked entries derived by three methods and describe scalability of the algorithms with respect to the number of observed entries in a tensor.

\begin{figure}[!htp]
	\includegraphics[width=0.45\textwidth]{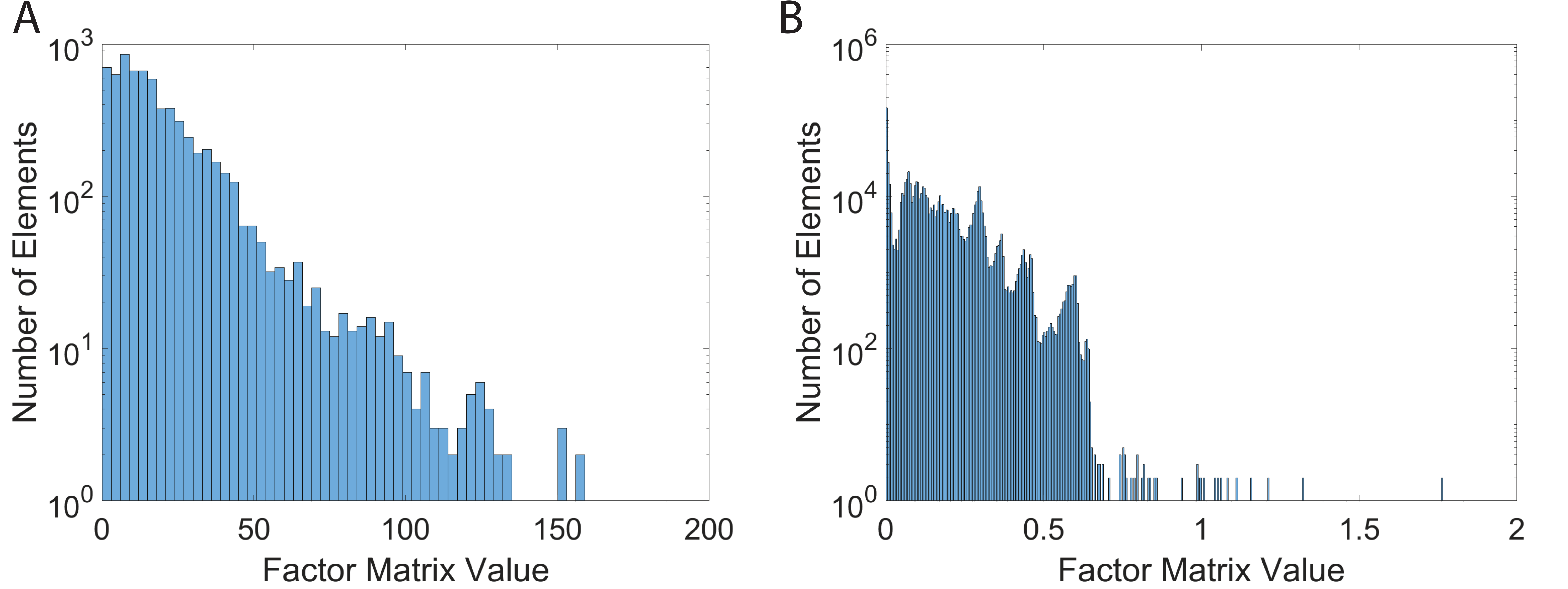}
	\caption{Distributions of values in a gene factor matrix derived by \gift ($\lambda$ = 100) for unmasked (\textbf{A}) and masked entries (\textbf{B}). The values of unmasked entries are much larger than that of masked ones. }
	\label{figS:exp:distribution1}
\end{figure}

Figure~\ref{figS:exp:distribution1} illustrates value distributions of masked and unmasked entries derived by \gift ($\lambda=100$). Compared to the result offered by the main paper ($\lambda=10$), the gap between masked and unmasked ones becomes much larger. Although it provides more interpretable results, its accuracy is lower than that of the case when $\lambda=10$. Moreover, it is hard to reveal important masked entries when $\lambda=100$. Thus, we use $\lambda=10$ in the discovery section.

\begin{figure}[!htp]
	\includegraphics[width=0.45\textwidth]{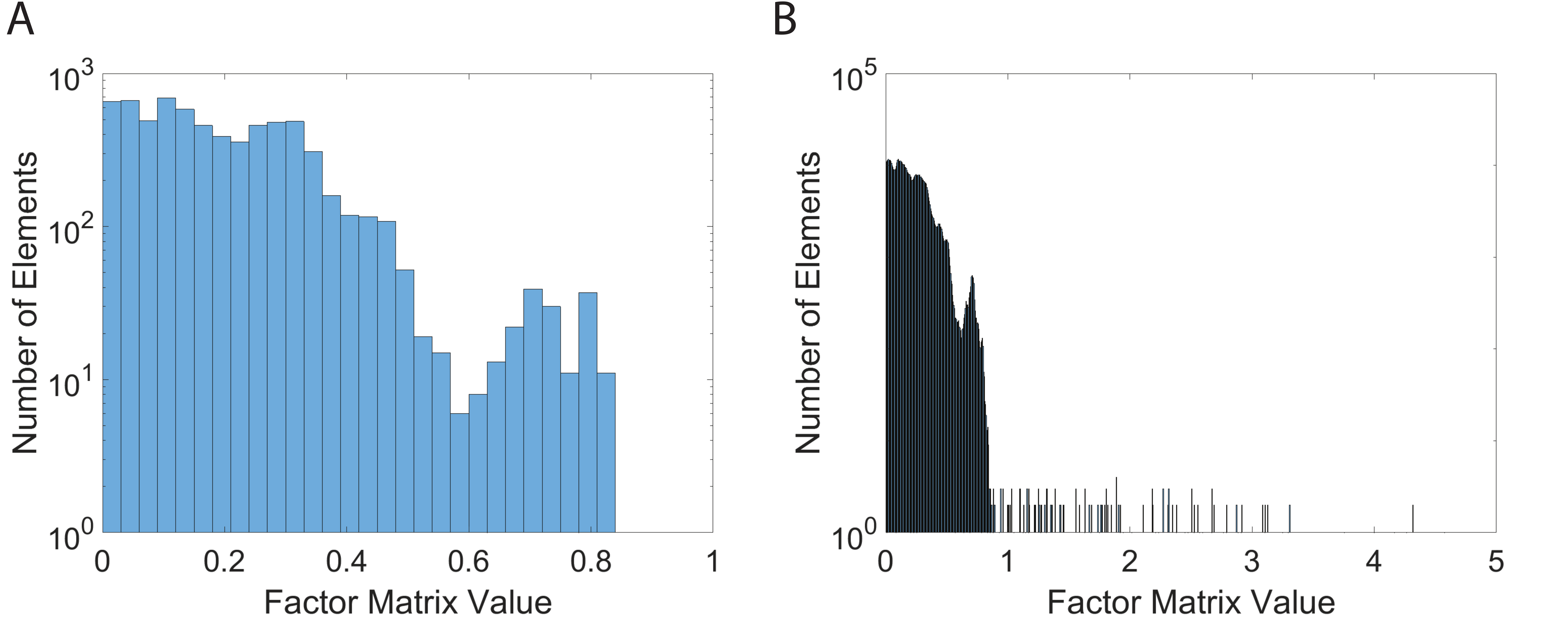}	\caption{Distributions of values in a gene factor matrix derived by \ptf ($\lambda$ = 10)for unmasked (\textbf{A}) and masked entries (\textbf{B}). There are no clear distinctions between unmasked and masked entries.}
	\label{figS:exp:distribution2}
\end{figure}

However, \ptf fails to make a distinction between masked and unmasked entries, as shown in Figure~\ref{figS:exp:distribution2}. The results are easily expected since \ptf does not differentiate the masked and unmasked ones when it updates factor matrices.

\begin{figure}[!htp]
	\includegraphics[width=0.45\textwidth]{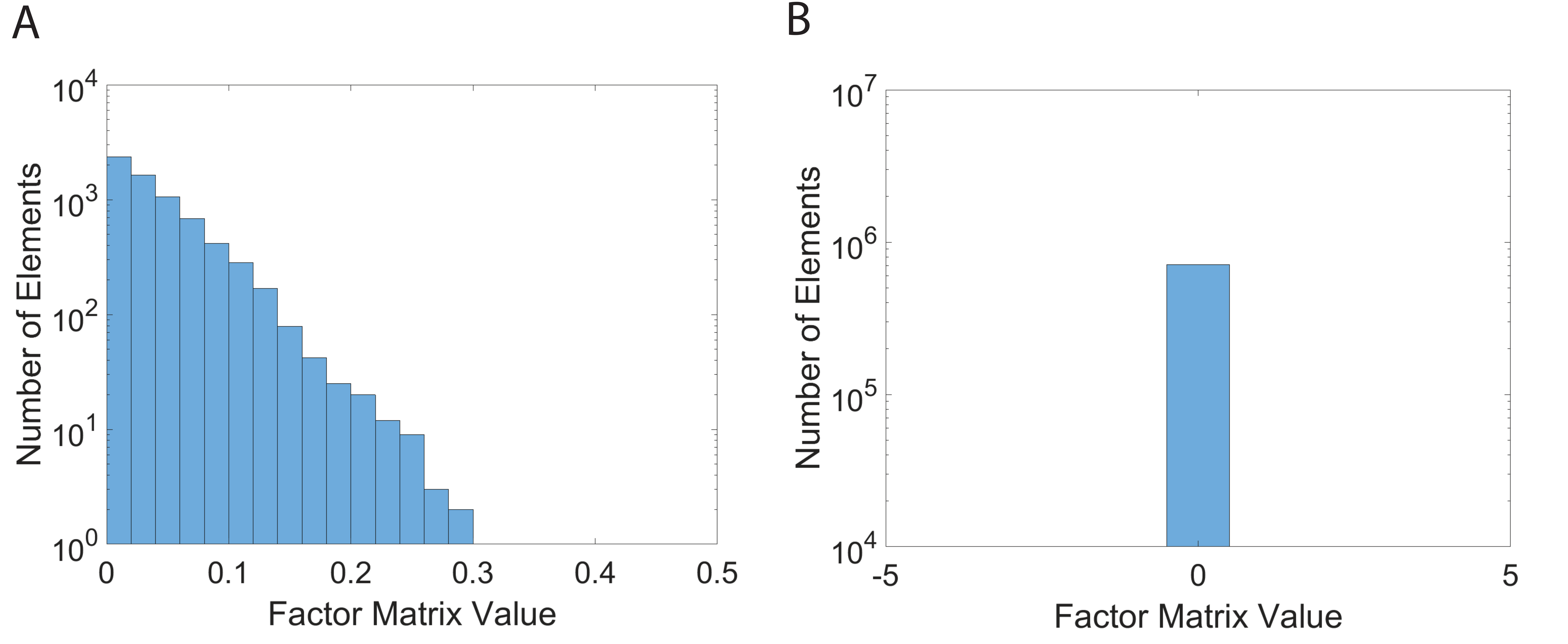}	\caption{Distributions of values in a gene factor matrix derived by \stf ($\lambda$ = 10) for unmasked (\textbf{A}) and masked entries (\textbf{B}). The values of masked entries are set to zeros, while the values of unmasked entries are varying from 0 to 0.3. }
	\label{figS:exp:distribution3}
\end{figure}

On the other hand, \stf produces interpretable results, as presented in Figure~\ref{figS:exp:distribution3}. The values of masked entries are fixed to zeros, while the values of unmasked entries are varying from 0 to 0.3. Although \stf provides interpretable results, it cannot retrieve important masked entries as the values of them are set to zeros.

\subsection{Scalability} \label{secS:exp:scalability}
\begin{figure} [t!]
	\includegraphics[width=0.45\textwidth]{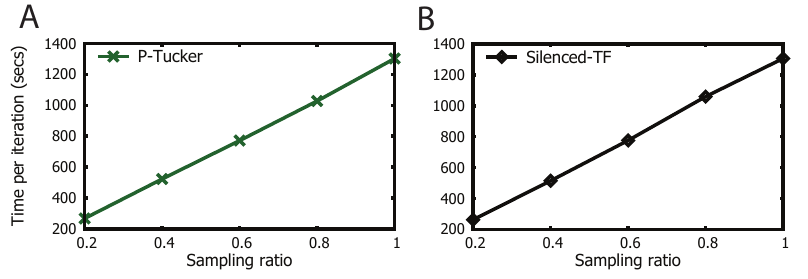}
	\centering
	\caption{Scalability of \ptf (\textbf{A}) and \stf (\textbf{B}) with respect to the number of observable entries in the tensor. As the number of observed entries increases, a running time of \ptf and \stf increases proportionally.}
	\label{figS:exp:scalability}
\end{figure}

We vary the number of observable entries by randomly sampling 20\%, 40\%, 60\%, 80\%, and 100\% from the PANCAN12 tensor. As shown in Figures~\ref{figS:exp:scalability}, \ptf (\textbf{A}) and \stf (\textbf{B}) scale near linearly in terms of the number of observable entries.

\end{document}